\documentclass[conference,10pt]{IEEEtran}
\usepackage[]{graphicx}
\usepackage[]{amsfonts,amsmath, amssymb,calrsfs}
\usepackage[mathscr]{eucal}

\DeclareMathAlphabet{\mathpzc}{OT1}{pzc}{m}{it}

\def\ket#1{| #1 \rangle}

\def\cC{\mathcal{C}}

\def\cH{\mathcal{H}}

\def\cN{\mathcal{N}}
\def\cP{\mathcal{P}}

\def\cS{\mathcal{S}}

\DeclareMathAlphabet{\mathpzc}{OT1}{pzc}{m}{it}

\def\eq#1{Eq.~\eqref{eq:#1}}
\newcommand{\argmax}{\operatornamewithlimits{argmax}}

\begin{document}

\title{Degenerate Viterbi Decoding}

\author{\authorblockN{Emilie Pelchat and David Poulin}
\authorblockA{
D\'epartement de Physique, Universit\'e de Sherbrooke, Sherbrooke, Qu\'ebec, Canada, J1K 2R1}}

\maketitle

\begin{abstract}
We present a decoding algorithm for quantum convolutional codes that finds the class of degenerate errors with the largest probability conditioned on a given error syndrome. The algorithm runs in time linear with the number of qubits. Previous decoding algorithms for quantum convolutional codes optimized the probability over individual errors instead of classes of degenerate errors. Using Monte Carlo simulations, we show that this modification to the decoding algorithm results in a significantly lower block error rate.
\end{abstract}


\IEEEpeerreviewmaketitle

\section{Introduction}

The stabilizer formalism \cite{Got97a} for quantum error correction reveals many similarities between classical and quantum codes. In particular, it shows that quantum codes can be obtained from classical self-dual codes over $\mathbb F_4$ \cite{CRSS98a}, and inherit many of their properties. Despite this connection, there exists fundamental distinctions between classical and quantum codes, among which degeneracy stands out \cite{SS96a,DSS98a,SS07b}.  Whereas two distinct bit-flip patterns $e$ and $e'$ applied to a given bit string $s$ always produces two distinct strings $s+e$ and $s+e'$, it is possible in quantum mechanics to find two distinct errors $E$ and $E'$ that have exactly the same effect on every code state, i.e. $E\ket{\overline \psi} = E'\ket{\overline \psi}$ for every state $\ket{\overline \psi}$ in the code space $\cC$. We say that errors $E$ and $E'$ are degenerate for the code $\cC$. In fact, every quantum code has degenerate errors, but we say that a code is degenerate when it has degenerate errors that are typical.\footnote{Thus, the notion of degeneracy makes implicit reference to an error model.} 

Degeneracy is a purely quantum mechanical effect with no classical counterpart, and leads to important distinctions between classical and quantum information theory. In particular, Shannon's bound to the capacity of a memoryless channel $\cN$
\begin{equation}
C(\cN) \leq \max_X I(X:\cN(X))
\end{equation}
is established by counting the number channel outputs that are produced by applying typical errors on all possible input codewords, and demanding that it does not exceed the total number of strings. If we apply the same counting argument to a degenerate quantum code, because some errors have the same effect on all codewords, we overestimate the number of distinct channel outputs and hence underestimate the channel capacity. Indeed, the quantum analog of the mutual information $I(X:\cN(X))$ for an input distribution $X$ and channel $\cN$ is the coherent information $I^c(I\otimes\cN(\phi_{AB}))$ for a purification $\phi_{AB}$ of an input distribution $\phi_B$ and a quantum channel $\cN$. Based on this analogy, one would expect that the quantity
\begin{equation}
Q_1(\cN) = \max_{\phi_B} I^c(I\otimes\cN(\phi_{AB}))
\end{equation}
upper bounds the quantum channel capacity. But due to the existence of degenerate errors, the true quantum capacity is instead given by regularized quantity \cite{Llo97a,Sho02a,Dev05a}
\begin{equation}
Q(\cN) = \lim_{n\rightarrow \infty}\frac 1n \max_{\phi_{B_n}} I^c(I\otimes\cN^{\otimes n}(\phi_{A^{n}B^{n}})).
\end{equation}
The quantity $Q_1$ can be achieved by non-degenerate codes---for Pauli noise defined below it is achieved by random stabilizer codes similar to those used in Shannon's original construction in the classical setting. However, $Q$ requires the use of degenerate codes. Because $Q$ involves an optimization over an input distribution correlated across many channel uses, it cannot be computed in general. However, degenerate codes of finite rate have been tailored for channels with $Q_1 = 0$ \cite{SS96a,DSS98a,SS07b}, a striking demonstration that $Q$ can be greater than $Q_1$. 

To benefit from the degeneracy of a quantum code, it is necessary to take it into consideration in the decoding process. Maximum a posteriori (MAP) decoding usually consists in identifying the error with largest probability conditioned on a given error syndrome. But since degenerate errors have the same effect on all code states, they can all be corrected the same way. Hence, for a degenerate quantum code, MAP decoding should instead consist of identifying the class of degenerate errors with the largest probability conditioned on a given error syndrome, the probability of a class of errors being the sum of the probabilities of its elements. This can be substantially more complicated than standard MAP decoding. Topological codes \cite{DKLP02a,Kit03a} are a good example of codes for which a standard MAP decoder exists \cite{DKLP02a} but no degenerate MAP decoder is known, and where we know that degeneracy can provide an advantage \cite{SB09a}. Thus, the design of degenerate MAP decoding algorithms is an important problem for quantum information theory.

In this article, we study this problem for quantum convolutional codes. MAP decoding of classical convolutional codes can be formally classified as a MAX-PROD problem, and is solved using Viterbi's algorithm \cite{Vit67a}. The MAX is to optimize the conditional probability over all errors, while the PROD reflects the fact that the probability of a given error on a memoryless channel is the product of the probability of each of its components. Viterbi's algorithm makes use of the distributive law $\max_{x,y} x\cdot y = \max_x x \cdot \max_y y$ to solve the MAP decoding problem \cite{AM00a} in a time that scales linearly with the length of the code. Using the stabilizer formalism, all this machinery can be imported to the quantum realm and yields a MAP decoder for quantum convolutional codes \cite{OT05a,PTO09a}. 

A degenerate MAP decoder has a different formal structure however; the need to add-up the probabilities of all degenerate errors entails a MAX-SUM-PROD problem. By making use of the distributive law of the product over the sum and of the max over the product, together with the particular factorized structure of the problem, we conceive a generalization of Viterbi's algorithm that exactly achieves degenerate MAP decoding of quantum convolutional code in linear time. Using Monte Carlo simulations, we find that the degenerate decoder suppresses more errors than the standard non-degenerate decoder; the improvement becoming more important at low error rates. Thus, convolutional codes provide an example where degenerate MAP decoding can be performed efficiently and yield a significant performance gain. 

The rest of this article is organized as follows. The next section introduces basic concepts. In Sec. \ref{sec:decoding}, we define the decoding problem for quantum codes, emphasizing on the role of degenerate errors. Section \ref{sec:viterbi} summarizes Viterbi's algorithm for quantum convolutional codes, and explains how it can be modified to take degeneracy into consideration. Lastly, we present numerical result in Sec. \ref{sec:results}. Our presentation of quantum convolutional codes and their decoding algorithms follows \cite{PTO09a}, we refer the reader to this article for more technical details on these basic concepts.

\section{Definitions}

\subsection{Stabilizer codes}
\label{sec:stab}

A quantum state of $n$ qubits is specified by a vector $\ket\psi$ in a $2^n$-dimensional vector space $\cH$. A quantum code $\cC$ encoding $k$ qubits into $n$ qubits is a $2^k$-dimensional subspace of $\cH$.  The stabilizer formalism offers a compact description of such a subspace, making use of the Pauli group. The Pauli matrices are defined as follows
\begin{equation}
X = \begin{pmatrix}0 & 1 \\ 1 & 0  \end{pmatrix}, \;\; Y = \begin{pmatrix}0 & -i \\ i & 0  \end{pmatrix},\;\;
Z = \begin{pmatrix}1 & 0 \\ 0 & -1  \end{pmatrix}.
\end{equation}
Together with the $2\times 2$ identity matrix $I$ and the imaginary unit $i$, they form a group under multiplication, the Pauli group $\cP_1$. The $n$-qubit Pauli group is the $n$-fold tensor product of the single-qubit Pauli group $\cP_n = \cP_1^{\otimes n}$. An important fact about $\cP_n$ is that all of its elements either commute or anti-commute, i.e. for all $P,Q \in \cP_n$, $PQ = \pm QP$. Using the fact that $ZX = iY$, we find that $\{i,X_a,Z_a\}$ is a generating set of $\cP_n$, where $Z_a$ ($X_a$) stands for the Pauli operator $Z$ ($X$) acting on qubit $a$, and is the identity elsewhere, i.e. 
\begin{equation}
Z_a = \underbrace{I\otimes I\otimes\ldots I}_{a-1}\otimes Z \otimes \underbrace{I\otimes I\otimes\ldots I}_{n-a-1}.
\end{equation}  

A stabilizer group $\cS$ is an Abelian subgroup of $\cP_n$ that does not contain the element $-I$. It can be specified by $s\leq n$ independent stabilizer generators $\{S_a\}_{a=1,\ldots,s}$. The quantum code $\cC$ associated to the stabilizer group $\cS$ is defined by a set of eigenvalue equations
\begin{equation}
\cC = \{ \ket{\overline \psi} : S\ket{\overline \psi} = \ket{\overline \psi}, \forall S\in \cS\}.
\label{eq:stab}
\end{equation}
Note that the condition $S_a\ket{\overline \psi} = \ket{\overline \psi}$ for all generators $S_a$ of $\cS$ is enough to ensure \eq{stab}. The dimension of the code $\cC$ is $2^k$ with $k=n-s$, so we say that the code encodes $k$ logical qubits into $n$ physical qubits, so has rate $\frac kn$. Clearly, the stabilizer generators $S_a$ play a role analogous to the linearly independent rows of the parity check matrix of a classical linear code. 

The centralizer $C(\cS)$ of $\cS$---i.e. elements of $\cP_n$ that commute with all elements of $\cS$---are called Pauli codewords. Multiplying a code state $\ket{\overline \psi} \in \cC$ by a Pauli codeword $L \in C(\cS)$ produces another code state. This can be verified directly since $S(L\ket{\overline \psi})   = L(S\ket{\overline \psi}) = L\ket{\overline \psi}$ which shows that $L\ket{\overline \psi}$ is an eigenstate of eigenvalue $+1$ of all stabilizers $S$, where we have used the commutativity of $L$ and $S$. More generally, all code states can be generated starting from an arbitrary fiducial code state $\ket{\psi_0}$ and multiplying it by a logical operator $L \in {\rm Alg}(C(\cS))$ in the algebra generated by the Pauli codewords. 

The encoding circuit is another method to specify a stabilizer code. The main advantages of this circuit-based definition is that all commutation constraints discussed above are automatically satisfied, and it is more suited for the definition of quantum convolutional codes. The Clifford group on $n$ qubits is the normalizer of the Pauli group in $U(2^n)$. In other words, Clifford transformations consist of unitary transformation $U$ on $n$ qubits for which $UPU^\dagger$ is a Pauli operator, for all Pauli operators $P$. Any Clifford transformation on $n$ qubits can be decomposed into a quantum circuit composed of controlled-not gates, Hadamard gates $H = \frac1{\sqrt 2}(\begin{smallmatrix} 1&1\\ 1&-1 \end{smallmatrix})$, and phase gates $R = (\begin{smallmatrix} 1&0\\ 0&i \end{smallmatrix})$.
Given an $n$-qubit Clifford transformation $U$, we can construct a stabilizer code with stabilizer generators 
\begin{equation}
S_a = UZ_aU^\dagger, \quad a = 1,\ldots, s
\label{eq:stab}
\end{equation}
and Pauli codewords
\begin{equation} 
\overline X_a = U X_{s+a}U^\dagger, \quad
\overline Z_a = U Z_{s+a}U^\dagger, \quad a = 1,\ldots,k.
\label{eq:logical}
\end{equation}
 For later use, it is also convenient to define operators that we call pure errors as
\begin{equation}
T_a = UX_aU^\dagger, \quad a = 1,\ldots, s.
\label{eq:pureerror}
\end{equation}

Since conjugation by a unitary matrix preserves the commutation relations, the following commutation relations are a straightforward consequence of definitions Eqs.~\ref{eq:stab}-\ref{eq:pureerror}:
\begin{align}
[S_a,S_b] &= 0 \label{eq:stab_comm} \\ 
[S_a,\overline Z_b] &=  [S_a,\overline X_b] = 0 \label{eq:log_comm} \\ 
S_aT_b &= (-1)^{\delta_{ab}} T_bS_a \label{eq:err_comm}
\end{align}
where \eq{stab_comm} expresses the fact that $\cS$ is Abelian, \eq{log_comm} shows that Pauli codewords are in the center of $\cS$, and \eq{err_comm} will be helpful in formulating the decoding problem. Also, because the operators $\{Z_a, X_a\}$ generate $\cP_n$ and because completeness is not affected by a unitary transformation, it follows that $\{i,\overline Z_a, \overline X_a, S_b,T_b\}_{a=1,\ldots, k, b=1,\ldots,s}$ generate $\cP_n$. 
Lastly, $U$ is called the encoding circuit because any state obtained from the following circuit
\begin{center}\includegraphics[scale=0.5]{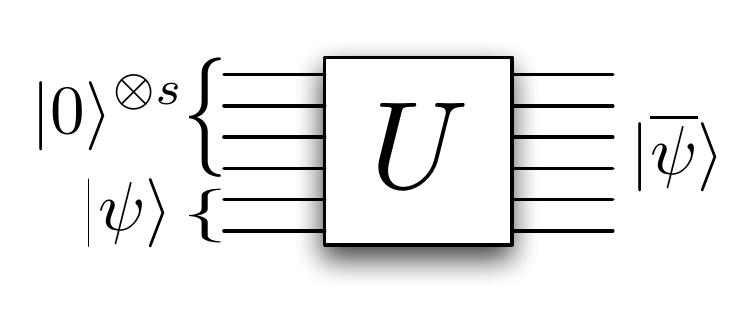}\end{center}
where $\ket\psi$ can be an arbitrary $k$-qubit state, is a code state, i.e. $\ket{\overline \psi} = U(\ket0^{\otimes s}\otimes \ket\psi) \in \cC$. In this circuit, we call the $s$ top input qubits the ancillary qubits, while the bottom $k$ input qubits are called the data qubits as they carry the information prior to encoding into $\cC$.

\subsection{Pauli noise}
\label{sec:noise}

The natural noise models to study using stabilizer codes are Pauli noise models. In these models, the noise alters an encoded state $\ket{\overline \psi}$ by multiplying it by an element of the Pauli group $E\in \cP_n$, i.e. $\ket{\overline \psi} \rightarrow E\ket{\overline \psi}$. The errors $E$ are chosen at random according to some probability $P(E)$ that specifies the noise model. A common assumption is that of an i.i.d. noise model where $P(E)$ is the $n$-fold tensor product of a distribution $(1-p_x-p_y-p_z, p_x,p_y,p_z)$ over the single qubit Pauli group.\footnote{Note that the absolute phase of the noise---e.g. whether error $X$, or $iX$, or $-$X was applied---has no observable effect, so our description of errors is over equivalent classes of Pauli operators defined modulo a phase.} The symmetric choice $p_x=p_y=p_z = p$ is called the depolarization channel of rate $p$. 

\subsection{Quantum convolutional codes}

Quantum convolutional codes were introduced in \cite{OT03a,OT04a}, but here we follow essentially the definitions of \cite{PTO09a}, and refer the reader to this article for more technical definitions. An $(n,k,\eta)$-quantum convolutional code is a stabilizer code with stabilizer generators of the form
\begin{equation}
\{S_{a,t} = I^{\otimes t\times n} \otimes S_a \}_{a=1,\ldots ,s; t=1,\ldots,T}
\label{eq:conv_stab}
\end{equation}
where $S_a \in \cP_{n+\eta}$ and $\eta$ is called the constraint length of the code.  In other words, there are $s$ distinct $(n+\eta)$-qubit Pauli operators $S_a$ that are translated by integer multiples of $n$ to generate the entire set of generators. Note that the total number of qubits used in the code is left unspecified in this definition; the maximum value $\tau$ of $t$ determines the length of the convolutional code. Thus, it is implicitly assumed in \eq{conv_stab} that the operators are padded to the right with identity matrices so that they all act on $N = n\tau+\eta$ qubits. Of course, the $S_a$ should all commute with one another to ensure $[S_{a,t},S_{a',t}]=0$, and also with their translations by integer multiples of $n$ to ensure $[S_{a,t},S_{a',t'}]=0$. For codes with large constraint length $\eta$, this last commutation condition can lead to a large number of constraints on the $S_a$ that are difficult to fulfill. But as for stabilizer codes discussed above, there exists an equivalent circuit-based definition of quantum convolutional codes that circumvents this difficulty. 

In this circuit-based definition, a $(n,k,m)$-quantum convolutional code is a stabilizer code whose encoding circuit takes the particular form
\begin{center}\includegraphics[scale=0.5]{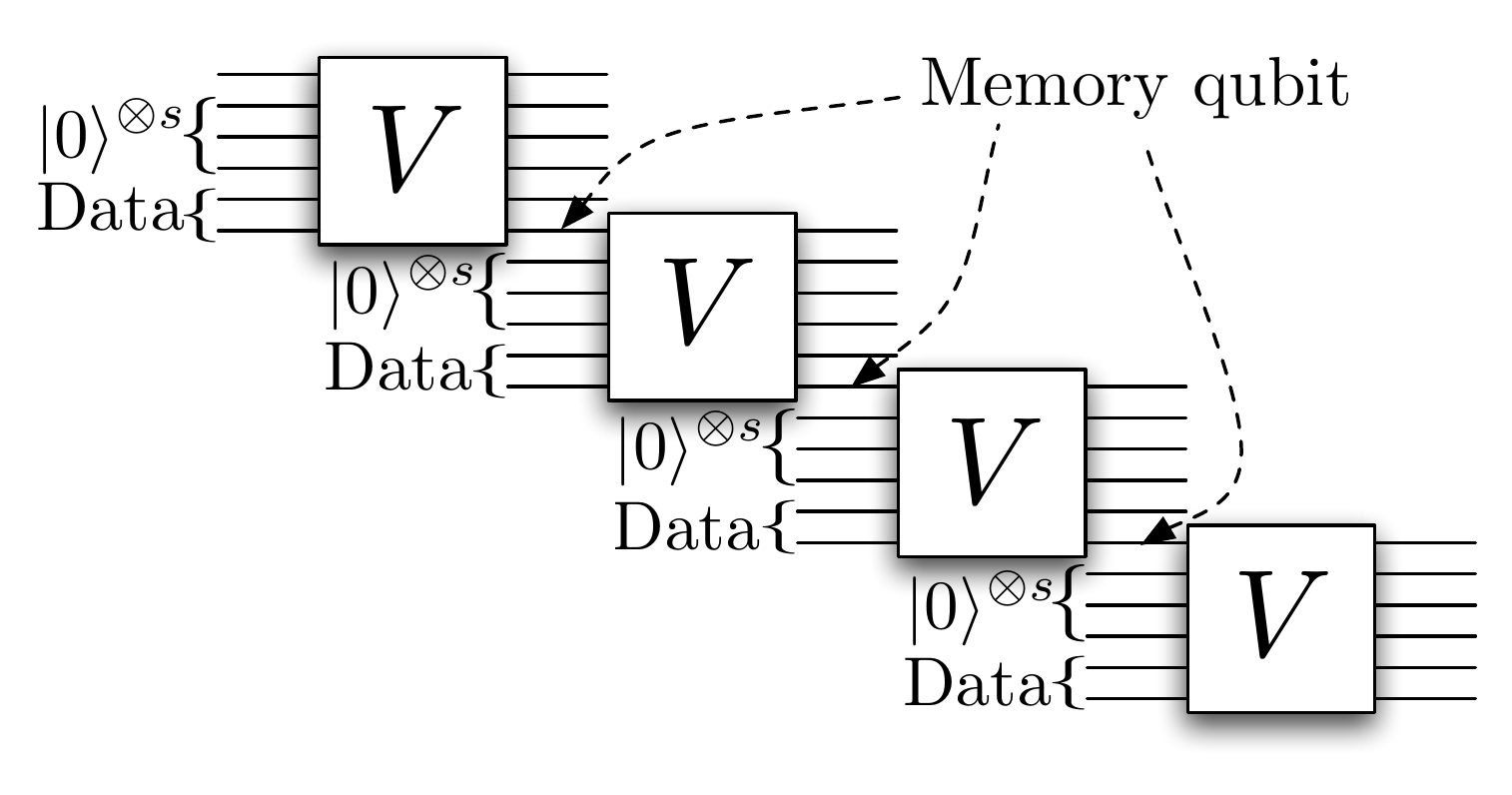}\end{center}
which illustrates the case $n=5$, $k=2$ and $m=1$. The encoding circuit is composed of a seed transformation $V$, element of the Clifford group, that is applied periodically to non-disjoint sets of qubits. We refer to a given period of this circuit as a time frame of the code.  The qubits that overlap two consecutive applications of the seed transformation are referred to as memory qubits, and there are in general $m$ of them in every frame. In each frame (except the very first one), there are a total of $n$ input qubits, $k$ of which carry quantum information (marked ``Data" in the circuit) and $s=n-k$ are ancillary qubits in the sate $\ket 0$. The rate of the code is therefore $\frac kn$.  A quantum convolutional code can contain an arbitrary number $\tau$ of frames. For a given value of $\tau$, we obtain a code block containing a total of $N = \tau \times n$ physical qubits and encoding $\tau \times k$ logical qubits.\footnote{Depending on the way the code is terminated, there can be a constant number of extra physical qubits in the block, which decreases the code's rate by a quantity that vanishes as $\frac 1\tau$, see \cite{FGG07a,PTO09a}.}

A generating set for the stabilizer is given by $S_a = U Z_a U^\dagger$ where $U$ is the $N$-qubit Clifford transformation resulting from the entire sequence of seed transformation $V$, and the qubit index $a$ varies over all ancillary qubit index, i.e. locations on the encoding circuit marked with qubits in a $\ket 0$ input state. 

The notation used in the definition based on \eq{conv_stab} and the definition based on the encoding circuit differ slightly because there is no direct relation between the constraint length $\eta$ and the number of memory qubits $m$. However, one can easily show that a convolutional code defined by a circuit with $m$ memory qubits admits a set of generators of constraint length $\eta \leq 4^m$.

\section{Degenerate decoding}
\label{sec:decoding}

As in the classical setting, the first step in the decoding process is the error syndrome extraction. In quantum mechanics, this is done by measuring the stabilizer generators. Remember that in quantum mechanics, the measurement of an observable---i.e. an Hermitian matrix---gives an outcome equal to the eigenvalue of the operator corresponding to the state of the system. By definition, all code states have $+1$ eigenvalues for all stabilizer generators, i.e. $S_a\ket{\overline \psi} = \ket{\overline \psi}$. If an error $E$ corrupts the state to $E\ket{\overline \psi}$, the eigenvalue of $S_a$ will remain the same when $E$ and $S_a$ commute, and will be changed to $-1$ when $E$ and $S_a$ anti commute. To see this, note that
\begin{equation*}
S_a(E\ket{\overline \psi}) = \left\{\begin{array}{ll}
ES_a\ket{\overline \psi} = E\ket{\overline \psi} & {\rm if\ } S_aE = ES_a \\
-ES_a\ket{\overline \psi} = -E\ket{\overline \psi} & {\rm \ if\ } S_aE = -ES_a.
\end{array}\right.
\end{equation*}
The $\pm 1$ measurement outcome of $S_a$ gives the $a$th syndrome bit $s_a$, and we denote the collection of $s$ syndrome bits ${\bf s} = (s_1,s_2,\ldots s_s)$.

The goal of the non-degenerate (ND) decoder can be stated quite simply: find the most probable error conditioned on the measured syndrome. If we denote ${\bf s}(E) = (s_1(E),\ldots,s_s(E))$ the syndrome that error $E$ would produce, non-degenerate MAP decoding consists of the optimization problem
\begin{equation}
E_{\rm MAP}^{\rm ND}({\bf s}) = \argmax_{E: {\bf s}(E) = {\bf s}} P(E)
\label{eq:MAP_ND}
\end{equation}
where $P(E)$ is given by the noise model, and argmax denotes the argument that achieves the maximum. 

To explain the degenerate decoding problem, it is convenient to express Pauli errors in a basis tailored to the stabilizer code. As explained in Sec.~\ref{sec:stab}, the set $\{i,\overline Z_a, \overline X_a, S_b,T_b\}$ forms a basis for $\cP_n$. Thus, an error $E$ has a unique decomposition into a product of logical operators, elements of the stabilizer group, pure errors, and an irrelevant phase factor that we will henceforth ignore. In other words, we can uniquely decompose any error as $E = \overline LST$ where $\overline L \in C(\cS)$, $S\in \cS$, and $T \in \langle T_a \rangle$. We can therefore interpret the noise model $P(E)$ as a probability distribution over $L$, $S$, and $T$, simply setting
\begin{equation}
P(\overline L,S,T) = P(E=\overline LST).
\end{equation}

Observe that the syndrome $\bf s$ is in one-to-one correspondence with the pure error component $T$, namely, an error with syndrome $\bf s$ has the pure error component
\begin{equation}
T({\bf s}) = \prod_a T_a^{(1-s_a)/2}
\end{equation}
where $T_a^0 = I$ and $T_a^1 = T_a$. This can be seen from the commutation relation \eq{err_comm}, which shows that $S_a$ anti-commutes with $E$ if and only if $E$ contains $T_a$ when decomposed in the basis $\{\overline Z_a, \overline X_a, S_b,T_b\}$. In addition, Eqs.~\ref{eq:stab_comm}-\ref{eq:log_comm} show that the $S$ and $L$ components of an error $E$ have no effect on its syndrome. Thus, knowledge of the error syndrome is equivalent to knowledge of $T$, so the error probability conditioned on the error syndrome is obtained by Bayes' rule
\begin{equation}
P(E|{\bf s}) = P(\overline L,S|T({\bf s})) = \frac{P(\overline L,S,T({\bf s}))}{P(T({\bf s}))}
\label{eq:conditional}
\end{equation}
where the marginal probability is defined as usual $P(T({\bf s})) = \sum_{\overline L,S} P(\overline L,S,T({\bf s}))$. 

Degenerate errors are those that differ only by their $S$ component. Indeed, the errors $E$ and $E' = ES$ for $S\in \cS$ have exactly the same effect on all code states by definition. Thus, only the logical component $\overline L$ needs to be identified to correct the error; the $T$ component is known given the syndrome and the $S$ component has no effect on the encoded information. The degenerate MAP decoding therefore  consists of the optimization problem 
\begin{equation}
L_{\rm MAP}^{\rm D}({\bf s}) = \argmax_{\overline L} P(\overline L|T({\bf s})),
\label{eq:MAP_D}
\end{equation}
where $ P(\overline L|T({\bf s})) = \sum_S  P(\overline L,S|T({\bf s}))$ is the marginal conditional distribution obtained from \eq{conditional}. 

An equivalent way to explain the degenerate decoding problem is to imagine un-encoding the corrupted encoded state $E \ket{\overline \psi}$. This produces the state  $U^\dagger E \ket{\overline \psi}  = U^\dagger E U (\ket\psi \otimes  \ket 0^{\otimes s}) = L\ket\psi \otimes Q \ket 0^{\otimes s}$ where $L\otimes I^{\otimes s} = U^\dagger \overline L U$ is the un-encoded version of the logical component $\overline L$ of $E$ and $I^{\otimes k}\otimes Q = U^\dagger TSU$ is the un-encoded version of the stabilizer $S$ and pure error $T$ components of $E$. Note that the un-encoded version of $S$ contains only $Z$ operators on the ancillary qubits by definition, c.f. \eq{stab}. Since $Z\ket 0 = \ket 0$, we conclude that the $S$ component has no effect on the un-encoded corrupted state, a direct manifestation of degeneracy. The $T$ component on the other hand map to $X$ operators under $U^\dagger$, c.f. \eq{pureerror}. We conclude that the final state is $U^\dagger E\ket{\overline \psi} = L\ket\psi \otimes \ket{\bf s}$ where $\ket{\bf s}$ is a shorthand for a $s$-qubit state with qubit $a$ in a state $\ket{\frac{1-s_a}2}$. The error syndrome can be directly obtained by measuring the ancillary qubits, and the degenerate decoding problem consists in identifying the most likely $L$ given $\bf s$. 

\section{Degenerate Viterbi algorithm}
\label{sec:viterbi}

\begin{figure}
\includegraphics[scale=0.5]{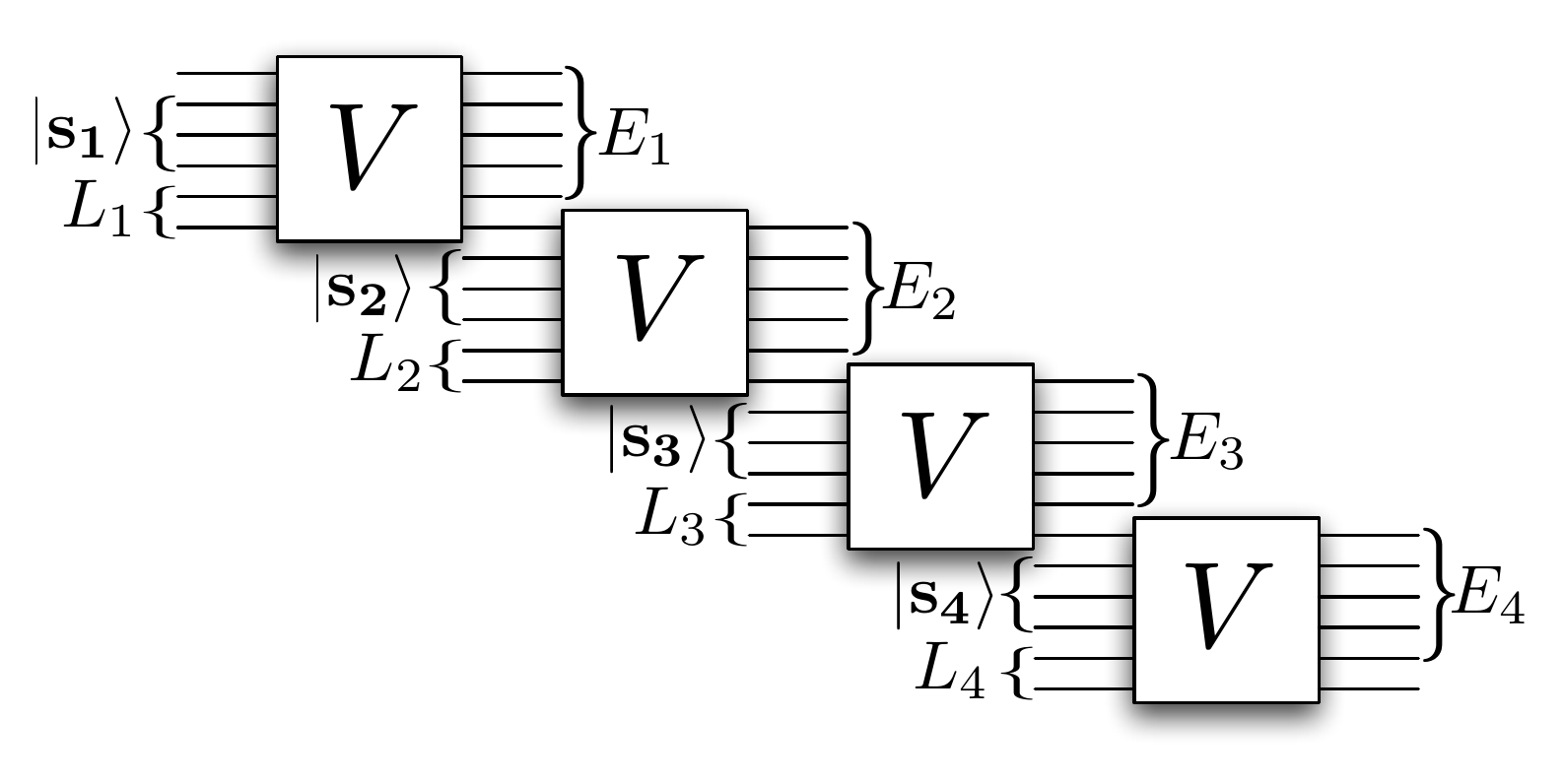}
\caption{Decomposition of an error $E$ into a tensor product of errors $E_t$ in each $n$-qubit frame. The error produces a syndrome $\bf s$ and an un-encoded logical component $L$, that are also broken-up into components on each frame ${\bf s}_t$ and $L_t$.  }
\label{fig:convE}
\end{figure}

With convolutional code, it is natural to use a notation that reflects the periodic structure of the encoding circuit. We use a two-index notation to identify qubits: qubit $(t,a)$ corresponds to the $a$th qubit of the $t$th time frame, in other words the $[(t-1)n+a]$th qubit altogether since there are $n$ qubit per time frame. Similarly, the error syndrome $\bf s$ naturally breaks into $\tau$ distinct $s$-bit syndromes ${\bf s} = ({\bf s}_1,\ldots ,{\bf s}_\tau)$ with  ${\bf s}_t = (s_{t,1},\ldots s_{t,s})$.  Following the last paragraph of the previous section, syndrome bit $s_{t,a}$ is obtained by measuring the ancillary qubit $(t,a)$ after having un-encoded the corrupted state, see Figure \ref{fig:convE}. Similarly, it is natural to also break $L$ and $E$ into a product of operators on each frame $L_t$ and $E_t$, see Figure \ref{fig:convE}.  Since we assume that the quantum channel is memoryless, the error probability factors as $P(E) = \prod_t P(E_t)$.

The standard, non-degenerate, decoding of quantum convolutional code uses a trellis. Trellis-based decoding of quantum convolutional codes was introduced in \cite{OT05a}, but here we follow the presentation of \cite{PTO09a}, and refer the reader to this article for more details of the construction. The trellis for an $(n,k,m)$-quantum convolutional code of length $\tau$ is a directed multigraph whose vertices can be grouped into $\tau$ sets $\Lambda_t$. Each element of $\Lambda_t$ is labeled by a distinct element $M_t \in \cP_m$, so $|\Lambda_t| = 4^m$. Given an error syndrome $\bf s$,  there is an edge between $M_{t-1} \in \Lambda_{t-1}$ and $M_{t} \in \Lambda_{t}$ if and only if there exists an $L_t \in \cP_k$, $Z_t \in \langle I,Z\rangle^s$, and $E_{t} \in \cP_{n-m}$ such that
\begin{equation}
V(M_{t-1}\otimes Z_t X({\bf s}_t)\otimes L_t)V^\dagger = E_{t}\otimes M_{t}   ,
\label{eq:condition}
\end{equation}
where $X({\bf s}_t) = \prod_{a=1}^s X_a^{(1-s_{t,a})/2}$. There can in general be different choices of $E_t$ that fulfill this condition, and we label the multiple edges by the associate $E_t$. This condition is illustrated by the circuit
\begin{center}\includegraphics[scale=0.5]{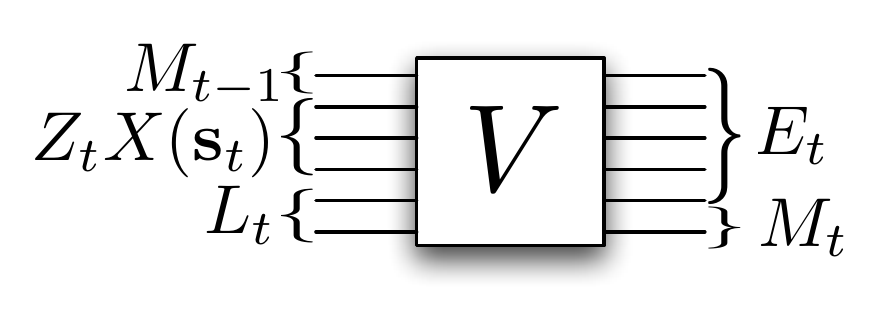}\end{center}
which can be understood, reading from right to left, as one segment of the un-encoding circuit which proceeds recursively starting at $t=\tau$ and making its way to $t=1$. The interpretation is a situation where errors in the previous un-encoding steps have resulted in a memory state $M_{t}$ which, combined with the error segment $E_{t}$, produces the right syndrome bits ${\bf s}_t$ for the time frame $t$, modifies the logical state by the application of $L_t$ in this frame, and sets the memory in a new state $M_{t-1}$. 

With this interpretation we see that paths in the trellis correspond to all the distinct errors $E$ with syndrome $\bf s$. To find the most likely error (or path) among them, we associate a probability (or weight) to each edge. The probability associated to the edge linking $M_{t-1}$ to $M_{t}$  labeled $E_t$ is $P(E_t)$, the probability of the corresponding error, specified by the error model. The non-degenerate MAP decoding then consists in finding the path in the trellis with the largest probability, where the probability of a path is equal to the product of the probability of each of its segments. It is convenient to speak instead of the length of a path, which is equal to the sum of the weight of its segments, the weight of segment $w(M_{t-1},M_t;E_t)$ labeled $E_t$ being $-\ln P(E_t)$, and we set $w(M_{t-1},M_t;E_t) = \infty$ if there is no edge labeled $E_t$ between $M_{t-1}$ and $M_t$. Then, the decoding problem becomes that of finding the shortest path in the trellis. 

Viterbi's algorithm solves this problem recursively, starting at $t=\tau$ and decreasing the value of $t$ by 1 at each iteration, by associating a cumulative distance to the vertices of the graph. The cumulative distance of vertex $M_t$, $d(M_t)$, is equal to the length of the shortest path starting at $t=\tau$ and leading to $M_t$. It obeys the recursive equation
\begin{equation}
d(M_{t-1}) = \min_{M_t,E_t} [d(M_t) + w(M_{t-1},M_t;E_t)],
\end{equation}
so the algorithm keeps only the edges that realize the maximum for each $M_{t-1}$, the other edges are erased. The initialization of the algorithm is given by $d(M_\tau) = -\ln P(M_\tau)$ where $P(M_\tau)$ is the error model on the last $m$ bits of the circuit. To explain the termination of the algorithm, we must specified that the first $m$ qubits of the quantum circuit (the memory qubits of frame 0) are used as additional ancillary qubits. Thus, they are initialized in state $\ket 0^{\otimes m}$ prior to encoding and measured in the $Z$ basis after the un-encoding, see \cite{PTO09a} for more details. The output of this measurement reveals the $X$ component of $M_0$, and only the values of $M_0$ with the correct component are kept. The shortest path is the one connecting to the $M_0$ with the shortest cummulative distance $d(M_0)$.  

The reason why the previous algorithm does not take degeneracy into account is that each path is associate to a distinct physical error. For degenerate decoding, errors need to be associated to equivalent classes of states instead, where errors that differ by an element of $\cS$ are joined in the same class. One way to keep track of such classes is to label them by their logical un-encoded component $L = L_1\otimes \ldots\otimes L_\tau$. The degenerate Viterbi algorithm will thus use the same trellis structure as above, but instead of labeling an edge that fulfills \eq{condition} by the corresponding $E_t$, it will be labeled by the corresponding un-encoded logical operator segments $L_t \in \cP_k$. As before, the probability associated to an edge will be equal to $P(E_t)$ for the $E_t$ fulfilling the condition \eq{condition}. Thus, at this point, the construction of the trellis follows exactly the original prescription of \cite{OT05a}, but its edges are labeled differently. 

\begin{figure}
\includegraphics[scale=0.42]{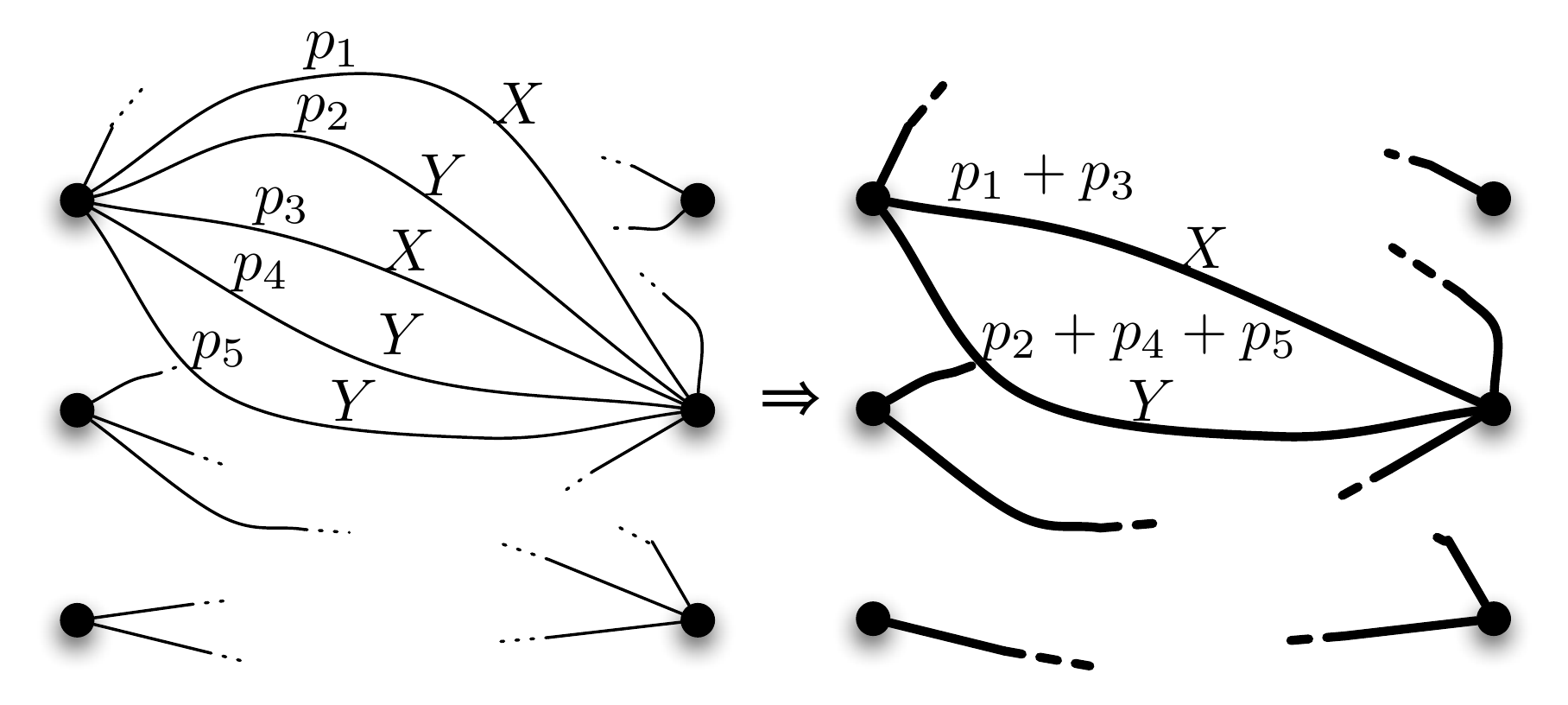}
\caption{Merging procedure.  Multiple edges with the same label linking $M_{t-1}$ to $M_t$ are merged into a unique super edge with the corresponding label. The probability of a super edge is equal to the sum of the probabilities of the merged edges. }
\label{fig:merge}
\end{figure}

For a given syndrome segment ${\bf s}_t$, un-encoded logical segment $L_t$, and memory states $M_{t-1}$ and $M_t$, there can be multiple distinct $E_t$ that fulfills the condition \eq{condition}. Call this set of solutions $\Omega(M_{t-1},M_t,L_t,{\rm s}_t)$. Thus, the trellis contains multiple edges between some memory states $M_{t-1}$ and $M_t$ with the same label $L_t$. These multiple edges $\Omega(M_{t-1},M_t,L_t,{\rm s}_t)$ correspond to degenerate paths, i.e. paths that differ only by an element of the stabilizer group. Therefore, they should be viewed as equivalent paths so we merge them into a single super edge as illustrated in Fig.~\ref{fig:merge}. The probability of a super edge is equal to the sum of the probabilities of the merged edges, or in terms of weight:
\begin{equation}
w(M_{t-1},M_t;L_t) = -\ln \left( \sum_{E_t\in \Omega(M_{t-1},M_t,L_t,{\rm s}_t)} P(E_t) \right).
\end{equation}
In terms of these weights, the degenerate decoding problem becomes the problem of finding the shortest path in the trellis,  for which Viterbi's algorithm can be used.

In summary, our modification to Viterbi's decoding algorithm consists in a pre-processing phase that modifies the trellis my merging edges that correspond to degenerate errors and computing the associated weights. Given this modified trellis, the degenerate decoding problem becomes identical to the non-degenerate decoding problem, and can be solved with Viterbi's original algorithm.

\section{Results}
\label{sec:results}

\begin{figure}
\includegraphics[width=9cm]{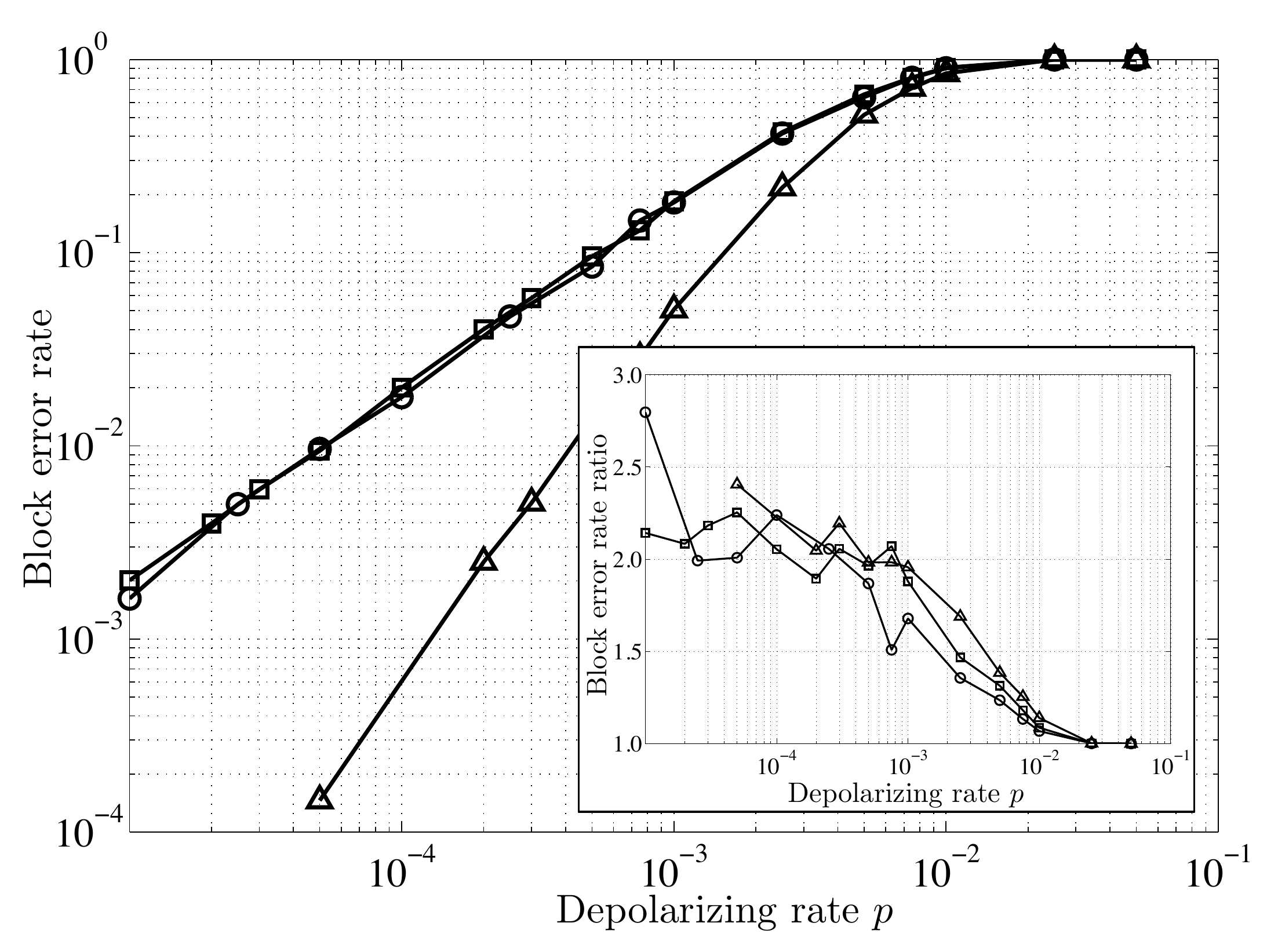}
\caption{Block error rate as a function of the channel depolarizing rate $p$ for different $(n,k,m)$-quantum convolutional codes: circles $(4,1,1)$; squares $(4,1,2)$; triangles $(5,1,3)$. Since all codes have $k=1$ and the length of the code is $\tau = 600$, all codes encode 600 logical qubits. The triangles represent a code of lower rate, so as expected it shows overall better performances.  Inset: the ratio of block error rates obtained with the non-degenerate Viterbi decoder and the degenerate Viterbi decoder. The  degenerate Viterbi decoder always yields a lower block error rate, and this advantage increases as the channel error rate decreases.}
\label{fig:results}
\end{figure}

We have benchmarked our degenerate Viterbi decoder using Monte Carlo simulations on quantum convolution codes of finite block length $\tau=600$ with various code parameters $(n,k,m)$, results are presented in Fig.~\ref{fig:results}. The codes were generated by choosing the seed transformation $V$ at random in the Clifford group. For each set of parameters $(n,k,m)$, we have simulated a few dozens of such randomly generated codes and kept the most interesting results, although our conclusions extend qualitatively to all codes we have simulated. The codes are terminated by padding, as explained in \cite{PTO09a}. 

The Monte Carlo simulations were realized as follows. An $N$ qubit error $E$ is generated randomly according the depolarizing error probability of rate $p$ (c.f. Sec.~\ref{sec:noise}) and the corresponding error syndrome $\bf s$ is calculated. The syndrome is fed as input to two distinct algorithms: a degenerate and a non-degenerate decoding algorithm. The degenerate decoder outputs the most likely logical error component $L_{\rm MAP}^{\rm D}({\bf s})$ given by \eq{MAP_D}. It is declared successful if the output $L_{\rm MAP}^{\rm D}({\bf s})$ belongs to the equivalence class of the randomly generated error $E$, and failed otherwise.  The procedure is repeated $N_{\rm sample}$ times  to accumulate statistics, and the block error rate  equals the frequency of its failures. The number of samples $N_{\rm sample}$ is adjusted such that at least 30 significant events are observed, resulting in a relative error at most $1/\sqrt{30}$ on the reported data. Figure \ref{fig:results} reports the block error rate of the degenerate decoder as a function of the channel error rate for different convolutional codes.

A similar procedure is performed in parallel with the non-degenerate Viterbi decoder during our Monte Carlo simulations. Given the input syndrome $\bf s$, the non-degenerate decoder outputs the most likely error $E_{\rm MAP}^{\rm ND}({\bf s})$ given by \eq{MAP_ND}.  It is declared successful if the output $E_{\rm MAP}^{\rm ND}({\bf s})$ belongs to the equivalence class of the randomly generated error $E$, and failed otherwise. Note that the non-degenerate decoder therefore benefit from the existence of degeneracy because its output $E_{\rm MAP}^{\rm ND}({\bf s})$ is not required to exactly match the randomly generated error $E$ to be accepted as a successful decoding, as long as they belong to the same equivalence class. The distinction is that degeneracy is not explicitly taken into account to estimate the error.

To evaluate the impact of degeneracy on the decoder's performance, we compare its block error rate with the one obtained with the non-degenerate Viterbi decoder. In the inset of Fig.~\ref{fig:results}, we show the ratio between the non-degenerate decoder's block error rate to the one of the degenerate decoder. The fact that all data points are above 1 indicates that the non-degenerate decoder always yields a lower block error rate than the non-degenerate decoder. We also observe in the inset of Fig.~\ref{fig:results} that the benefit becomes more prominent at low depolarizing rates. This effect appears to be independent of the code parameters: the block error rate ratios of all the codes we have simulated show a clear monotonic increase as $p$ decreases. We have observed gains as large as 4.4 dB. 

\section{Conclusion}

Degeneracy is a key feature of quantum codes that is at the origin of important distinctions between classical and quantum information theory. To gain from the existence of degeneracy, it must be taken into account during the decoding process of a code. Here, we have developed a decoding algorithm for quantum convolutional codes that exactly solves the maximum a posteriori decoding problem over equivalent classes of degenerate errors. The main modification to the standard Viterbi algorithm is a pre-processing phase that modifies the trellis. Our Monte Carlo simulations show that degenerate decoding improves the error suppression of the code, and that this effect becomes more prominent at low error rates.

Because convolutional codes are the main ingredient of quantum turbo codes, a natural next step would be to use our degenerate decoder in a concatenated scheme, where its benefit could be further amplified. The development of a fault-tolerant computing scheme based on these codes is also desirable. The importance of our decoder may be particularly important in this setting since fault-tolerant quantum computers operate at low error rates, where the effect of degenerate decoding are most prominent.

\medskip
\noindent {\em Acknowledgements---}This research was partially funded by Mprime, NSERC, FQRNT and the IARPA QCS program. Numerical resources were provided by Calcul Qu\'ebec and Compute Canada.

\noindent {\em Disclaimer---} The views and conclusions contained herein are those of
the authors and should not be interpreted as necessarily representing
the official policies or endorsements, either expressed or implied, of
IARPA, DoI/NBC, or the U.S. Government.

%\bibliographystyle{IEEEtran}
%\bibliography{qubib,code}
%\bibliographystyle{/Users/dpoulin/archive/hsiam}
%\bibliography{/Users/dpoulin/archive/qubib}

\begin{thebibliography}{10}
\providecommand{\url}[1]{#1}
\csname url@samestyle\endcsname
\providecommand{\newblock}{\relax}
\providecommand{\bibinfo}[2]{#2}
\providecommand{\BIBentrySTDinterwordspacing}{\spaceskip=0pt\relax}
\providecommand{\BIBentryALTinterwordstretchfactor}{4}
\providecommand{\BIBentryALTinterwordspacing}{\spaceskip=\fontdimen2\font plus
\BIBentryALTinterwordstretchfactor\fontdimen3\font minus
  \fontdimen4\font\relax}
\providecommand{\BIBforeignlanguage}[2]{{%
\expandafter\ifx\csname l@#1\endcsname\relax
\typeout{** WARNING: IEEEtran.bst: No hyphenation pattern has been}%
\typeout{** loaded for the language `#1'. Using the pattern for}%
\typeout{** the default language instead.}%
\else
\language=\csname l@#1\endcsname
\fi
#2}}
\providecommand{\BIBdecl}{\relax}
\BIBdecl

\bibitem{Got97a}
D.~Gottesman, ``Stabilizer codes and quantum error correction,'' Ph.D.
  dissertation, California Institute of Technology, Pasadena, CA, 1997.

\bibitem{CRSS98a}
A.~R. Calderbank, E.~M. Rains, P.~W. Shor, and N.~J.~A. Sloane, ``Quantum error
  correction via codes over {GF(4)},'' \emph{IEEE Trans. Info. Theor.},
  vol.~44, p. 1369, 1998.

\bibitem{SS96a}
P.~W. Shor and J.~A. Smolin, ``Quantum error-correcting codes need not
  completely reveal the error syndrome,'' AT{\&}T Research, Murray Hill, NJ
  07974, 1996, preprint.

\bibitem{DSS98a}
D.~P. DiVincenzo, P.~W. Shor, and J.~A. Smolin, ``Quantum-channel capacity of
  very noisy channels,'' \emph{Phys. Rev. A}, vol.~57, pp. 830--839, 1998.

\bibitem{SS07b}
G.~Smith and J.~A. Smolin, ``Degenerate coding for {P}auli channels,''
  \emph{Phys. Rev. Lett.}, vol.~98, p. 030501, 2007.

\bibitem{Llo97a}
S.~Lloyd, ``Capacity of the noisy quantum channel,'' \emph{Phys. Rev. A},
  vol.~55, p. 1613, 1997.

\bibitem{Sho02a}
P.~Shor, ``The quantum channel capacity and coherent information,'' 2002, mSRI
  Workshop on quantum computation.

\bibitem{Dev05a}
I.~Devetak, ``The private classical capacity and quantum capacity of a quantum
  channel,'' \emph{IEEE Trans. Info. Theor.}, vol.~51, p.~44, 2005.

\bibitem{DKLP02a}
E.~Dennis, A.~Kitaev, A.~Landahl, and J.~Preskill, ``Topological quantum
  memory,'' \emph{J. Math. Phys.}, vol.~43, p. 4452, 2002.

\bibitem{Kit03a}
A.~Y. Kitaev, ``Fault-tolerant quantum computation by anyons,'' \emph{Ann.
  Phys.}, vol. 303, p.~2, 2003.

\bibitem{SB09a}
T.~Stace and S.~Barrett, ``Error correction and degeneracy in surface codes
  suffering loss,'' 2009.

\bibitem{Vit67a}
A.~J. Viterbi, ``Error bounds for convolutional codes and an asymptotically
  optimum decoding algorithm,'' \emph{IEEE Trans. Info. Theor.}, vol.~13,
  no.~2, pp. 260--269, 1967.

\bibitem{AM00a}
S.~Aji and R.~Mc{E}liece, ``The generalized distributive law,'' \emph{IEEE
  Trans. Info. Theor.}, vol.~46, no.~2, p. 325, 2000.

\bibitem{OT05a}
H.~Ollivier and J.-P. Tillich, ``Trellises for stabilizer codes: definition and
  uses,'' \emph{Phys. Rev. A}, vol.~74, p. 032304, 2006.

\bibitem{PTO09a}
D.~Poulin, J.-P. Tillich, and H.~Ollivier, ``Quantum serial turbo-codes,''
  \emph{IEEE Trans. Info. Theor.}, vol.~55, no.~6, p. 2776, 2009.

\bibitem{OT03a}
H.~Ollivier and J.-P. Tillich, ``Description of a quantum convolutional code,''
  \emph{Phys. Rev. Lett.}, vol.~91, no.~17, p. 177902, 2003.

\bibitem{OT04a}
------, ``Quantum convolutional codes: fundamentals,'' 2004.

\bibitem{FGG07a}
J.~G.~D. Forney, M.~Grassl, and S.~Guha, ``Convolutional and tail-biting
  quantum error-correcting codes,'' \emph{IEEE Trans. Info. Theor.}, vol.~53,
  p. 865, 2007.

\end{thebibliography}
%\end{document}


\end{document}